%% file: main.tex
\documentclass{llncs}
\usepackage{gastex}

\usepackage[english]{babel}
\usepackage[latin1]{inputenc}
\usepackage[T1]{fontenc}
\usepackage{epsfig}
\usepackage{graphicx}
\usepackage{amssymb}
\usepackage{amsfonts}
\usepackage{amsmath}
\usepackage{verbatim}
\usepackage{url}
\usepackage{xcolor}
\usepackage{gastex}
\usepackage{floatflt}
\usepackage{stmaryrd}
\usepackage[all]{xypic}
\usepackage{epic,eepic}
\usepackage{color}
\usepackage{algorithmic}
\usepackage{algorithm}

\xdefinecolor{Red}{rgb}{1,0.2,0}
\xdefinecolor{Green}{rgb}{0,0.7,0}
\xdefinecolor{Yellow}{rgb}{1,1,0}
\xdefinecolor{White}{rgb}{1,1,1}
\xdefinecolor{Blue}{rgb}{0,0,1}
\xdefinecolor{Black}{rgb}{0,0,0}
\xdefinecolor{Orange}{rgb}{1,1,0}

\title{The Cost of Parameterized Reachability in Mobile Ad Hoc Networks}
\author{Giorgio Delzanno$^1$ \and Arnaud Sangnier$^2$ \and\\  Riccardo Traverso$^1$ \and Gianluigi Zavattaro$^3$}
\institute{$^1$ University of Genova, Italy \\ $^2$ LIAFA, Univ Paris Diderot, Paris Cit\'e Sorbonne, CNRS, France \\$^3$ University of Bologna, INRIA - FOCUS Research Team, Italy}
\input{macros}
\begin{document}
\maketitle
\begin{abstract}
We investigate the impact of spontaneous movement in the complexity of verification problems 
for an automata-based protocol model of networks with selective broadcast communication.
We first consider reachability of an error state and show that parameterized verification
is decidable with polynomial complexity. We then move to richer queries and show how the complexity changes
when considering properties with negation or cardinality constraints.
%The discuss related properties for networks with message, node, and link failures. 
\end{abstract}
\input intro
\input ahn

%\input mahn
\input reach_logic
\input pos_queries

\input neg_queries
\input full_queries_petri
\input reachalg
%\input full_queries
%input conflict
\input concl

\input biblio
%\appendix
%\newpage
%\input{proof-appendix}
\end{document}

%% file: macros.tex
\newcommand{\set}[1]{\{ #1 \} }
\newcommand{\tuple}[1]{\langle #1 \rangle}

\newcommand{\broadcast}[1]{!!{#1}} % stile Esparza, Finkel, Mayr -- Lics '99
\newcommand{\receive}[1]{??{#1}}

\newcommand{\nat}{\mathbb{N}}

\newcommand{\state}{L}
\newcommand{\conf}{\gamma}
\newcommand{\confs}{{\cal C}}

\newcommand{\sbp}{ {\mathcal{P}} }

\newcommand{\pdef}{\mathcal{P}}
\newcommand{\ahn}{\mathit{MAHN}}

\newcommand{\confszero}{\confs_0}

\newcommand{\npspace}{\textsc{NPSpace}}
\newcommand{\expspace}{\textsc{ExpSpace}}
\newcommand{\logspace}{\textsc{LogSpace}}

\newcommand{\ptime}{\textsc{PTime}}

\newcommand{\np}{\textsc{NP}}
\newcommand{\pspace}{\textsc{PSpace}}

\newcommand{\RQ}{CC}
\newcommand{\RQPos}{\RQ[$\geq 1$]}

\newcommand{\PRP}{PRP}
\newcommand{\PQRP}{PRP}
\newcommand{\postadd}{\mathtt{postAdd}}
\newcommand{\postdel}{\mathtt{postDel}}

%% file: intro.tex
\section{Introduction}
Selective broadcast communication is often used in networks in which individual nodes have no precise information about the underlying connection topology (e.g. ad hoc wireless networks). As shown in \cite{NH06,Mer09,MBS10,SCPJ08,Concur09,fossacs}, this type of communication can naturally be specified in models in which a network configuration is represented as a graph and in which individual nodes run an instance of a given protocol specification. 
A protocol typically specifies a sequence of control states in which a node can %either 
send a message (emitter role), wait for a message (receiver role), or perform an update of its internal state. Selective broadcast communication is modeled as a 
simultaneous update of the state of the emitter node and of the states of its neighbors.

Already at this level of abstraction, verification of protocols with selective broadcast communication turns out to be a very difficult task. A formal account of this problem is given in \cite{concur,fossacs}, where the {\em control state reachability problem} is proved to be undecidable in an automata-based protocol model in which configurations are arbitrary graphs. 
The control state reachability problem consists in verifying the existence of an initial network configuration (with unknown size and topology) that may evolve into a configuration in which at least one node is in a given control state.
If such a control state represents a protocol error, then this problem naturally expresses (the complement of) a safety verification task in a setting in which nodes have no information a priori about the size and connection topology of the underlying network. 

In presence of spontaneous movement, i.e., non-deterministic reconfigurations of the network during an execution, control state 
reachability becomes decidable \cite{concur}.  
In this paper we focus on the complexity of different types of parameterized reachability problems in presence of spontaneous movement.
More precisely, we consider reachability queries defined over assertions that: (PRP) check the 
presence or absence of control states in a given configuration generated by an initial configuration of arbitrary size, and (CRP) cardinality queries that check the exact number of occurrences of control states in a reachable configuration (the counterpart of classical reachability).
The first and the second problem require, at least in principle, the exploration of 
an infinite-state space. Indeed they are formulated for arbitrary initial configurations.
The latter is inherently finite-state. 
Despite of it, we first show that reachability queries for constraints that only check for the presence of a control state can be checked in polynomial time.
When considering both constraints for checking presence and absence of control states the problem turns out to be \np-complete.
Finally, we show that the problem becomes \pspace-complete for cardinality queries.

\mbox{}\smallskip\\
{\bf Related Work.}
Perfect synchronous semantics for broadcast communication have been proposed in \cite{Pra95,SCPJ08,Concur09,EM01}.
Semantics that take into consideration interferences and conflicts during a transmission
have been proposed in \cite{G07,Mer09,MBS10,MS06}.
%In all these approaches a broadcast communication is split into several phases to model scenarios in which different transmission periods of different emitters overlaps. In this paper we also consider a semantics for explicitly representing conflicts. Differently from other models, our semantics allows multiple nodes to start a communication in the same instant. In \cite{concur,fossacs} we have studied decision problems for verification of models of ad hoc networks with selective broadcast communication with perfect semantics and no conflicts.
To our knowledge, parameterized verification has not been studied in previous work on formal models of ad hoc networks \cite{Pra95,SCPJ08,Concur09,NH06,EM01,FVM07,G07,Mer09,MBS10,MS06}. 
Finally, decidability issues for broadcast communication in unstructured concurrent systems have been studied, e.g., in \cite{EFM99}, whereas verification of unreliable communicating FIFO systems have been studied, e.g., in \cite{ic96}. 

%% file: ahn.tex
\newcommand{\iconfs}{{\confs^{int}}}
\newcommand{\iconfszero}{{\confs^{int}_{0}}}
\newcommand{\istate}{\state}
\newcommand{\inttrel}{\rightarrow}
\newcommand{\rightintr}{\dashleftarrow}
\newcommand{\downintr}{\dashrightarrow}

\newcommand{\deriv}{\mapsto}
\newcommand{\derivv}{\rightarrow}
\newcommand{\decomp}[1]{dec(#1)}
\newcommand{\ghostb}[1]{#1_{b}}
\newcommand{\ghostr}[1]{#1_{r}}
\newcommand{\broad}[1]{#1\!\!\uparrow}
\newcommand{\sem}[1]{[\!\![#1]\!\!]}
\newcommand{\semm}[2]{[\!\![\!\![#1,#2]\!\!]\!\!]}

\newcommand{\activated}{\mathtt{A}}
\newcommand{\deactivated}{\mathtt{D}}

\section{A Model for Mobile Ad Hoc Networks}
\label{sec:model}
\subsection{Syntax and semantics}
Our model for mobile ad hoc networks is defined in two steps. We first define graphs used to denote network 
configurations and then define protocols running on each node. The label of a node denotes its current control state.
Finally, we give a transition system for describing the interaction of a vicinity during the execution of 
the same protocol on each node.
\begin{definition}
A $Q$-graph is a labeled undirected graph $\gamma=\tuple{V,E,L}$, where $V$ is a finite set of {\em nodes},
$E\subseteq V\times V$ is a finite set of {\em edges}, and $L$ is a labeling function from $V$ to a %finite
set of labels $Q$.
\end{definition}
We use $L(\gamma)$ to represent all the labels present in $\gamma$ (i.e. the image of the function $L$).
The nodes belonging to an edge are called the {\em endpoints} of the edge.
For an edge $\tuple{u,v}$ in $E$, we use the notation $u\sim_{\gamma} v$ and say that the vertices $u$ and $v$ are adjacent one to another in the graph $\gamma$. We omit $\gamma$, and simply write
$u\sim v$, when it is made clear by the context.
\begin{definition}
A process is a tuple $\pdef=\tuple{Q,\Sigma,R,Q_0}$, where $Q$ is a finite set of control states, $\Sigma$ is a finite alphabet, $R \subseteq Q \times (\set{\broadcast{a}, \receive{a} \mid a  \in \Sigma}) \times Q$ is the transition relation, and $Q_0 \subseteq Q$ is a set of initial control states.
\end{definition}
% Arn 120212
%The label $\mytau$ represents the capability of performing an internal action,
%and 
The label $\broadcast{a}$ [resp. $\receive{a}$] represents the capability of broadcasting [resp. receiving] a message $a\in\Sigma$. 
For $q\in Q$ and $a \in \Sigma$, we define the set $R_{a}(q)=\set{q'\in Q \mid \tuple{q,\receive{a},q'} \in R}$ which
contains the states that can be reached from the state $q$ when receiving the message $a$.
We assume that $R_{a}(q)$ is non empty for every $a$ and $q$, i.e. nodes always react to broadcast messages.
Local transitions (denoted by the special label $\tau$) can be derived by using a special message $m_\tau$
such that $\tuple{q,\receive{m_\tau},q'}$ implies $q'=q$ for every $q \in Q$ (i.e. receivers do not modify their local states). In the following, if for some 
state $q \in Q$ and message $a \in \Sigma$ we omit the definition
of transitions of the form $\tuple{q,\receive{a},q'}$, we implicitly
assume the existence of only one such transition that does not
change the state (i.e. $q'=q$).

%Notice that rendezvous cannot be directly simulated using broadcast communication.
 
Given a process $\pdef=\tuple{Q,\Sigma,R,Q_0}$, a configuration of the corresponding Mobile Ad Hoc Network (MAHN)  is a $Q$-graph and an initial configuration is a $Q_0$-graph. We use $\confs$ [resp. $\confszero$] to denote  the set of configurations [resp. initial configurations] associated  to $\pdef$. Note that even if $Q$ is finite, there are infinitely many possible configurations (the number of $Q$-graphs).
We assume that each node of the graph is a process that runs a common predefined protocol defined by a communicating automaton with a finite set $Q$ of control states.
Communication is achieved via selective broadcast, which means that a broadcasted message is received by the nodes which are adjacent to the sender. We next formalize the above intuition.

%\begin{definition}
%Given a process $\pdef=\tuple{Q,\Sigma,R,Q_0}$, a configuration is a $Q$-graph and an initial configuration is a 
%$Q_0$-graph.
%\end{definition}
 Given a process $\pdef= \tuple{Q,\Sigma,R,Q_0}$, a MAHN is defined by the transition system $\ahn(\pdef)=\tuple{\confs,\inttrel,\confszero}$ where the transition relation $\inttrel \subseteq \confs \times \confs$ is such that: for $\gamma,\gamma' \in \confs$ with $\gamma=\tuple{V,E,\istate}$, we have 
$\gamma\inttrel\gamma'$ iff $\gamma'=\tuple{V,E',\istate'}$ 
and one of the following conditions holds:
\begin{description}
\item[Broadcast] $E'=E$ and $\exists v \in V$ s.t.  $\tuple{\istate(v),\broadcast{a},\istate'(v)}\in R$ and 
                  $\istate'(u)\in R_a(\istate(u))$ for every $u\sim v$, 
                  and $\istate(w)=\istate'(w)$ for any other node $w$.
\item[Movement] $E'\subseteq V\times V$ and $\istate = \istate'$.
\end{description}
We use $\inttrel^\ast$ to denote the reflexive and transitive closure of $\inttrel$.
%An execution path $\pi$ in $\ahn(\pdef)$ is a sequence $\gamma_0\gamma_1\ldots$ such that $\gamma_0 \in \confs_0$ and $\gamma_i\inttrel\gamma_{i+1}$ for all $i\geq 0$.
%We use $\pi[i]$ to denote the $i$-th configuration of path $\pi$ and we denote by $Path(\gamma)$ the set of execution paths that start in $\gamma$.

%% file: reach_logic.tex
%
 %the same as \PRP 

\subsection{Parameterized Reachability Problems}
A natural class of verification problems for MAHN consists in determining whether there exists an initial configuration from which a configuration respecting some constraints can be reached. 
%Gio 19/2
%This problem is a parameterized problem, since the number of nodes and the connectivity graph of the initial configuration is not fixed a priori. 
In this work, the constraints are boolean combination of atoms which allow to state the presence or the absence of a control state in a configuration. Given a process $\sbp=\tuple{Q,\Sigma,R,Q_0}$, a \emph{constraint}  over $\sbp$ is defined by the following grammar: $\varphi ::= \# q \geq 1~|~\# q=0 ~|~ \varphi\wedge \varphi~|~\varphi\vee \varphi$ with $q\in Q$. We denote by \RQ\ the class of constraints and by \RQPos\ the class of constraints  in which atomic propositions have only the form $\# q\geq 1$ (there exists at least one occurrence of $q$).
Given a configuration $\gamma$ the satisfaction relation $\models$ for constraints is defined by (we omit boolean cases defined as usual): $\conf \models \#q\geq 1$ iff $q \in L(\gamma)$ and $\conf \models \#q=0$ iff $q \notin L(\gamma)$.

%
% Arn 150212
%\begin{itemize}
%\item $\conf\models \#q\sim k$ iff $\Occ_q(\conf) \sim k$;
%\item $\conf\models \varphi\wedge\psi$ iff $\conf\models \varphi$ and $\conf\models\psi$;
%\item $\conf\models \varphi\vee\psi$ iff $\conf\models \varphi$ or $\conf\models\psi$;
%\item $\conf\models \neg \varphi$ iff $\conf\not\models \varphi$.

%\item $\conf\models \mathbf{EF}~\varphi$ iff there exists $\pi\in Path(\conf)$ such that $\pi[i]\models \varphi$ for some $i\geq 0$;
%\end{itemize}
%
% Arn120212
%
%In the rest of the paper we often use the abbreviation $q$ to denote the atom $q\geq 1$ (i.e. the presence of 
%at least one occurrence of  label $q$) and $\overline{q}$ to denote the atom $q=0$ (absence of label $q$).

The parameterized reachability problem (\PRP) can then be stated as follows:
\begin{description}
\item[Input:] A process $\sbp$ with  $\ahn(\sbp)=\tuple{\confs,\inttrel,\confs_0}$ and a constraint $\varphi$.
%\item[Input:] A process $\sbp=\tuple{Q,\Sigma,R,Q_0}$ with  $\ahn(\sbp)=\tuple{\confs,\Rightarrow,\confs_0}$, 
%              and a query $\varphi$.
\item[Output:] Yes, if $\exists\gamma_0 \in \confs_0$ and $\gamma_1\in\confs$ s.t. $\gamma_0\inttrel^*\gamma_1$ and $\gamma_1 \models\varphi$.
\end{description}
If the answer to this problem is yes, we will write $\sbp \models \Diamond \varphi$.
%Gio 19/2
We use the term {\em parameterized} to remark that the initial configuration is not fixed a priori.
In fact, the only constraint that we put on the initial configuration is that the nodes have labels taken from
$Q_0$ without any information on their number or connection links.
As a special case we can define the control state reachability problems studied in \cite{concur} 
as the \PQRP\ for the simple constraint $\# q \geq 1$ (i.e. is there an initial configuration that can reach 
a configuration in which the state $q$ is exposed?). 

We also remark that according to the semantics, the number of nodes stays constant in each execution starting from the same initial configuration.  As a consequence, when fixing the initial configuration  $\gamma_0$, we obtain finitely many possible reachable configurations.  Thus, checking if there exists $\gamma_1$ reachable from a given $\gamma_0$ s.t. 
$\gamma_1\models\varphi$ for a constraint $\varphi$ is a decidable problem.

On the other hand, checking the parameterized version of the reachability problem is in general more difficult. Indeed, in \cite{concur}, it is proved that \PQRP\ for simple constraints of the form  $\#q\geq 1$  is undecidable when deleting the movement rule from the semantics 
(i.e. nodes communicate via selective broadcast but the connectivity graph never changes during a computation). 
In \cite{concur}, it is also proved that \PQRP\ for the same class of simple constraints is decidable. However, the proposed decidability proof is based on a reduction to the problem of coverability in Petri nets which is known to be \expspace-complete \cite{rackoff78,lipton76}. Since no lower-bound was provided,  the precise complexity of \PRP\ with simple constraints was left as an open
problem that we close in this paper by showing that it is \ptime-complete.

%% file: pos_queries.tex
\section{\PRP\ restricted to constraints in \RQPos\ }
In this section, we study \PRP\ restricted to \RQPos. Note that this class of constraints allow to characterize configurations in which a given set of control states is present but they cannot express neither the absence of states nor  the number of their occurrences.
%
% Arn 150212
%
%In this section, we study \PRP\ restricted to \RQPos. This class of constraints allows to define reachability questions for which the target set of states (which correspond 
%correspond to upward closed set of configurations (we only require that a given set of labels occurs in a configuration).
We first give a lower bound for this problem.
\begin{proposition}\label{theorem:ptime}
\PRP\ restricted to \RQPos\ is \ptime-hard.
\end{proposition}
\begin{proof}
The proof is based on a \logspace-reduction from the Circuit Value Problem (CVP) which is know to be \ptime-complete \cite{Lad77}.
CVP is defined as follows: given an acyclic Boolean circuit with  $k$ input variables, $m$ boolean gates (of type and, or, not), a single output variable and  a truth assignment for the input variables, is the value of the output equal to a given boolean value?

Assume an instance of CVP $C$ with input/output/intermediate value names taken from a finite set $VN$. 
We denote by $v_{1},\ldots,v_{k} \in VN$ the inputs and by $v \in VN$ the output. 
Furthermore, each gate $g$ is represented by its signature $g(\odot,i_1,i_2,o)$ with $i_1,i_2,o\in VN$ and
$\odot\in\{\vee,\wedge\}$ or by $g(\neg,i,o)$ with $i,o\in VN$. Finally, let $b_1,\ldots,b_k \in \{true, false\}$ be a truth assignment for the inputs and $b \in \{true, false\}$ the value for the output to be tested.

%The encoding is based on two types of node.
The process $\pdef_C$ associated to $C$ has two types of initial states: $q_0$ (init nodes),
and $g$ (gate nodes) for each gate $g$ of $C$.
%Furthermore, we also need a constant  number of auxiliary states for each gate $g$.
%
A node in state $q_0$ broadcasts (an arbitrary number of) messages that model the  initial assignments to input variables.
Since the assignment is fixed, broadcasting these messages several times (or receiving them from different initial nodes) 
does not harm the correctness of the encoding.
When receiving an evaluation for their inputs (from an initial node or another gate node), a gate node evaluates the corresponding boolean function and then repeatedly broadcasts the value of the corresponding output.
Since $C$ is acyclic, once computed, the output value remains always the same (i.e. recomputing it does not harm).
Finally, reception of a value $v$ for output $z$ sends a $q_0$ node into state $ok$.
Reachability of an output value $v$ reduces then to \PRP\ for the process $\pdef_C$ with $ok$  the control state to be reached.

Formally, the process rules are defined as follows.
For $i \in \{1,\ldots,k\}$, we have rules
$\tuple{q_0,\broadcast{(v_i=b_i)},\linebreak[0]q_0}$ and
$\tuple{q_0,\receive{(v=b)},ok}$.
They model the assignment of value $v_i$ to input $x_i$ and reception of output value $v$.
%Notice that the input values are fixed.

For gate $g(\odot,i_1,i_2,o)$ and for each assignment $\alpha=\tuple{b'_1,b'_2}$ 
(with $b'_1,b'_2 \in \{true,false\})$ of values to $\tuple{i_1,i_2}$
(a constant number for each gate),  we associate the following subprotocol:
\begin{center}
\scalebox{0.8}{
\begin{picture}(105,40)(-5,-5)
  \node(A)(0,15){$g$}
  \node(B)(30,0){$g_1^\alpha$}
  \node(C)(30,30){$g_2^\alpha$}
  \node(D)(60,15){$g_f^\alpha$}
  \drawedge(A,C){$\receive{(i_1=b'_1)}$}
  \drawedge(C,D){$\receive{(i_2=b'_2)}$}
  \drawedge[ELside=r](A,B){$\receive{(i_2=b'_2)}$}
  \drawedge[ELside=r](B,D){$\receive{(i_1=b'_1)}$}
  \drawloop[loopangle=0](D){$\broadcast{(o=b'_1\odot b'_2})$}
\end{picture}
}
\end{center}
(Self-loops associated to receptions for which there are no explicit rules are omitted). 
We use a similar encoding for a $not$ gate.

Consider now the resulting process $\pdef_{C}=\tuple{Q,\Sigma,R,\{q_0\}\cup\{g\ |\ g \text{ is a gate in } C\}}$
with corresponding  transition system $\ahn = \tuple{\confs,\inttrel,\confszero}$.
We have that there exists $\gamma \in \confszero$ and $\gamma'$ in $\confs$
s.t. $\gamma \inttrel^\ast \gamma'$ and $\gamma' \models \#  ok \geq 1$ 
iff $b$ is the value for $v$ in $C$ with input values $b_1,\ldots,b_k$.
%
% Arn 170212
%
%The above described reduction requires logarithmic space in the size of $C$ (we need variables $i_{1},i_{2},o$ ---or $i,o$ for a not gate--- to
%represent input and output variables of a gate, each one requiring $log |C|$ bits;
%for each gate we produce a constant number of rules).
\qed
\end{proof}
We now show that \PRP\ restricted to \RQPos\ is in \ptime. The main idea to obtain this result lies in the fact that we can compute in polynomial time the set of control states that appear in the reachable configurations. 
%
% Arn 150212
%
%The algorithm is based on the construction of a symbolic execution path in which set of configurations are abstractly represented by set of control states.
The construction is based on the following key points.
We first observe that, in order to decide if control state $q$ can be reached, 
we can focus our attention on initial complete graphs (i.e. graphs in which all pairs of nodes are connected).
Indeed, spontaneous movement can be applied to non-deterministically transform a topology into any other one.
Another key observation is that if a configuration $\gamma$ can be reached from an initial configuration $\gamma_0$, 
then for any natural $k$, there exists a complete graph which is reachable from an initial configuration $\gamma_0'$ 
and in which each of the control states appearing in $\gamma$ appears at least $k$ times. 
The initial configuration $\gamma_0'$ is obtained by replicating $k$-times the initial graph $\gamma_0$.
The replicated parts are then connected in all possible ways (to obtain a connected graph).
We can then use spontaneous movement to activate and deactivate the different subparts in order to mimick 
$k$ parallel executions of the original system. 
For what concerns constraints in \RQPos\, this property of \PRP\ avoids the need of counting the occurrences of states.
We just have to remember which states can be generated by repeatedly applying process rules.
\begin{algorithm}[htbp] \footnotesize
\caption{Computing the set of control states reachable in a MAHN}
\label{alg:sfr1}
\begin{algorithmic}
\REQUIRE $\pdef=\tuple{Q,\Sigma,R,Q_0}$ a process
\ENSURE  $S \subseteq Q $ the set of reachable control states in $\ahn(\sbp)$
\STATE $S:=Q_0$
\STATE  $\mathit{oldS}:=\emptyset$
\WHILE{$S \neq \mathit{oldS}$}
\STATE $\mathit{oldS}:=S$
\FORALL{ $\tuple{q_1, \broadcast{a}, q_2} \in R$ such that $q_1 \in \mathit{oldS}$}
\STATE $S:=S \cup \{q_2\} \cup \{q' \in Q \mid \tuple{q,\receive{a},q'}\in R \land q \in \mathit{oldS}\}$
\ENDFOR
\ENDWHILE
\end{algorithmic}
\end{algorithm}
%
% Arn 120212
%
%\begin{algorithm}[htbp] \footnotesize
%\caption{Computing the set of control states reachable in a MAHN}
%\label{alg:sfr1}
%\begin{algorithmic}
%    \REQUIRE $\pdef=\tuple{Q,\Sigma,R,Q_0}$ a process
%    \ENSURE $\mathcal{S} \subseteq Q $ the set of reachable control states in $\ahn(\sbp)$
%    \STATE $\mathcal{S}:=Q_0$
%    \STATE $old\mathcal{S}:=\emptyset$
%    \STATE $\mathit{Msg}:=\emptyset$
%    \WHILE{$\mathcal{S} \neq \mathit{old}\mathcal{S}$}
%        \STATE $\mathit{old}\mathcal{S}:=\mathcal{S}$
%        \FORALL{ $\tuple{q_1, e, q_2} \in R$ such that $q_1 \in \mathit{old}\mathcal{S}$}
%            \IF{$\exists a \in \Sigma$ such that $e =~\broadcast{a} \lor (e =~\receive{a} \land a \in Msg)$}
%                \STATE $\mathcal{S} := \mathcal{S} \cup \{q_2\}$
%                \STATE $Msg := Msg \cup \{a\}$
%            \ENDIF
%        \ENDFOR
%    \ENDWHILE
%\end{algorithmic} 
%\end{algorithm}
%
By exploiting the above mentioned observations, when defining the decision procedure for checking control state reachability we can take the following assumptions: $(i)$ forget about the topology underlying the initial configuration; $(ii)$ forget about the number of occurrences of control states in a configuration (if it is reached once, it can be reached an arbitrary number of times by considering larger initial configurations as explained before);
(iii) consider a single symbolic path in which at each step we apply all possible rules whose preconditions can be satisfied in the current set and then collect the resulting set of computed states.

We now formalize the previous observations. Let $\pdef=\tuple{Q,\Sigma,R,Q_0}$ be a process with $\ahn(\pdef)=\tuple{\confs,\inttrel,\confszero}$ and let $\mathtt{Reach}(\sbp)$ be the set of reachable control states equals to $\{q \in Q \mid \exists \gamma \in \confszero.\exists \gamma'  \in \confs. \mbox{ s.t. } \gamma \inttrel^\ast \gamma' \mbox{ and } q \in L(\gamma')\}$. We will now prove that Algorithm \ref{alg:sfr1} computes $\mathtt{Reach}(\sbp)$. Let $S$ be the result of the Algorithm \ref{alg:sfr1} (note that this algorithm necessarily terminates because the \textbf{while}-loop is performed at most $|Q|$ times). We have then the following lemma.
\begin{lemma}\label{lemma:algo} The two following properties hold:
\begin{description}
\item[(i)] There exist two configurations $\gamma_0 \in \confszero$ and $\gamma \in \confs$ such that $\gamma_0 \inttrel^\ast \gamma$ and $L(\gamma)=S$.
\item[(ii)] $S=\mathtt{Reach}(P)$.
\end{description}
\end{lemma}
\begin{proof}
We first prove (i). We denote by  $S_0, S_1, \ldots, S_n$ the content of $S$ after each iteration of the loop of the Algorithm \ref{alg:sfr1}. We recall  that a graph $\gamma=\tuple{V,E,L}$ is complete if $\tuple{v,v'} \in E$ or $\tuple{v',v} \in E$ for all $v,v' \in V$. We will now consider the following statement: for all $j \in \set{0,n}$, for all $k \in \nat$, there exists a complete graph $\gamma_{j,k}=\tuple{V,E,\istate}$ in $\confs$ verifying the two following points:
\begin{enumerate}
\item $L(\gamma_{j,k})=S_j$ and for each $q \in S_j$, the set $\set{v \in V \mid \state(v)=q}$ has more than $k$ elements (i.e. for each element $q$ of $S_j$ there are more than $k$ nodes in $\gamma_{j,k}$ labeled with $q$),
\item there exits $\gamma_0 \in \confszero$ such that $\gamma_0 \inttrel^\ast \gamma_{j,k}$.
\end{enumerate}
To prove this statement we reason by induction on $j$. First, for $j=0$, the property is true, because for each $k \in \nat$, the graph $\gamma_{0,k}$ corresponds to the complete graphs where each of the initial control states appears at least $k$ times. We now assume that the property is true for all naturals smaller than $j$ (with $j <n$) and we will show it is true for $j+1$. We consider $E$ the set $\set{\tuple{\tuple{q_1,\broadcast{a},q_2},\tuple{q,\receive{a},q'}} \in R \times R \mid q_1,q \in S_j}$ and and $M$ its cardinality. Let $k \in \nat$  and let $N=k+2*k*M$. We consider the graph $\gamma_{j,N}$ where each control state present in $S_j$ appears at least $N$ times (such a graph exists by the induction hypothesis). From $\gamma_{j,N}$, we build the graph $\gamma_{j+1,k}$ obtained by repeating $k$ times the following operations: 
\begin{itemize}
\item for each pair $\tuple{\tuple{q_1,\broadcast{a},q_2},\tuple{q,\receive{a},q'}} \in E$, select a node labeled by $q_1$ and one labeled by $q$ 
and update their label respectively to $q_2$ and $q'$ (this simulates a broadcast from the node labeled by $q_1$ received by the node labeled $q$ in the configuration in which all the other nodes have been disconnected thanks to the movement and reconnected after). Note that the two selected nodes can communicate because the graph is complete.
\end{itemize}
By applying these rules it is then clear that $\gamma_{j,N} \inttrel^\ast \gamma_{j+1,k}$ and also that $\gamma_{j+1,k}$ verifies the property $1$ of the
statement. Since by induction hypothesis, we have that there exists $\gamma_0 \in \confszero$ such that $\gamma_0 \inttrel^\ast \gamma_{j,N}$, we also deduce that $\gamma_0 \inttrel^\ast \gamma_{j+1,k}$, hence the property 2 of the statement also holds. From this we  deduce that (i) is true.

To prove (ii), from (i) we have that $S \subseteq \mathtt{Reach}(\sbp)$
and we now prove that $\mathtt{Reach}(\sbp) \subseteq S$. 
Let $q \in \mathtt{Reach}(\sbp)$. We show that $q \in S$ 
by induction on the minimal length of an
execution path $\gamma_0 \inttrel^\ast \gamma$ such that 
$\gamma_0 \in \confszero$
and $q \in L(\gamma)$.
If the length is $0$ then $q \in Q_0$ hence also $q \in S$.
Otherwise, let $\gamma' \inttrel \gamma$ be the
last transition of the execution.
We have that there exists $q_1 \in L(\gamma')$ such that
$\tuple{q_1, \broadcast{a}, q} \in R$ [or
$q_1,q_2 \in L(\gamma')$ such that $\tuple{q_1, \broadcast{a}, q_3},
\tuple{q_2, \receive{a}, q} \in R$]. By induction hypothesis 
we have that $q_1 \in S$ [or $q_1,q_2 \in S$].
By construction, we can conclude that also $q \in S$. 
%, from (i) we have that $S \subseteq \mathtt{Reach}(\sbp)$, and $\mathtt{Reach}(\sbp) \subseteq S$ follows directly from the definition of the transition relation $\inttrel$ and from the construction of the set $S$.  
\qed 
\end{proof}
Since constraints in \RQPos\ check only the presence of states
and do not contain negation, given a configuration $\gamma$
and a constraint $\varphi$ in \RQPos\
such that $\gamma \models \varphi$, we also have that 
$\gamma' \models \varphi$ for every $\gamma'$ such that 
$L(\gamma) \subseteq L(\gamma')$. Moreover, 
given a process $\sbp$,
by definition of $\mathtt{Reach}(\sbp)$ we have that
$L(\gamma)\subseteq \mathtt{Reach}(\sbp)$ for every
reachable configuration $\gamma$, and by Lemma \ref{lemma:algo}
there exists a reachable configuration $\gamma_f$ such that
$L(\gamma_f)=\mathtt{Reach}(\sbp)$.
Hence, to check $\sbp \models \Diamond \varphi$ it is sufficient 
to verify whether
$\gamma_f \models \varphi$ for such a configuration $\gamma_f$.
%This allows us to simply checks whether such configuration $\gamma'$
%satisfies $\varphi$.
%Hence,
This can be done algorithmically as follows:
once the set $\mathtt{Reach}(\sbp)$ is computed, 
check if the boolean formula obtained from $\varphi$ by replacing each atomic constraint of the form $\# q \geq 1$ by $\mathit{true}$ if $q \in \mathtt{Reach}(\sbp)$ and by $\mathit{false}$ otherwise is valid. 
%In fact, if this holds then from Lemma \ref{lemma:algo} {\bf (i)}
%we have that there exists a reachable configuration $\gamma$ such that 
%$L(\gamma)=S$ hence also $\gamma \models \varphi$.
%On the other hand, if there exists a reachable configuration $\gamma$
%such that $\gamma \models \varphi$, 
%by Lemma \ref{lemma:algo} {\bf (ii)} we have that 
%$L(\gamma) \subseteq \mathtt{Reach}(\sbp)$.
%Since $\varphi$ is in \RQPos, it is a ``monotone''
%formula that does not contain any negation,
%i.e. as it is satisfied by $\gamma$, it is also satisfied
%by any other ``greater'' configuration that includes
%at least the labeled nodes in $\gamma$.
%Hence, the boolean formula obtained from $\varphi$ as described
%above should also holds.
%deduce that the obtained formula is valid if and only if there exists a reachable configuration $\gamma$ such that $\gamma \models \varphi$ (we recall that since $\varphi$ is in \RQPos , it does not contain any negation).
This allows us to state the following theorem.
\begin{theorem}
\PRP\ restricted to \RQPos\ is \ptime-complete.
\end{theorem}
\begin{proof}
The lower bound is given by Proposition \ref{theorem:ptime}. 
To obtain the upper bound, it suffices to remark that the Algorithm \ref{alg:sfr1} is in \textsc{Ptime} since it requires at most $|Q|$ iterations
each one requiring at most $|R|^2$ look-ups (of active broadcast/receive transitions) for computing new states to be included, and also that evaluating the validity of a boolean formula can be done in polynomial time.  \qed
\end{proof}

%% file: neg_queries.tex
\section{Complexity for \PRP\ }
%
%In this section we study the decidability and complexity of \PRP\ restricted to cardinality constraints in \RQNeg. 
In this section we study the decidability and complexity of \PRP\ for  constraints in \RQ. 
The main difference with the problem studied in the previous section lies in the fact that now the constraints have the ability to specify that a given control state is not present in a configuration (using atomic constraints of the form $\# q =0$). Authorizing this kind of atomic constraints leads to a complexity jump as stated by the following proposition whose proof can be found in Appendix.
%
% Arn 170212
%
%We now study the decidability and complexity of the reachability problem for the subclass of formulas
%obtained by restricting atomic constraints to formulas of the form $\#q=0$ or $\# q\geq 1$ 
%(abbreviated as $\overline{q}$ and $q$).
%
\begin{proposition}\label{theorem:nphard} 
%\PRP\ restricted to \RQNeg\ is \np-hard.
\PRP\ for constraints in \RQ\ is \np-hard.
\end{proposition}
\begin{proof}
The proof is based on a reduction of the boolean satisfiability problem (SAT) which is known to be \np-complete.
Let $\Phi$ be a boolean formula in conjunctive normal form over the set of variables $V=\{v_1,\ldots,v_k\}$.
We define a process $\sbp$ with initial state $q_0$ and the following set of  rules 
$R=\{ \tuple{q_0,\tau,v}~|~v\in V\}\cup \{ \tuple{q_0,\tau,\overline{v}}~|~v\in V\}$.
%
% Arn 170212
%
%$$
%R=\{ \tuple{q_0,\tau,v}~|~v\in V\}\cup \{ \tuple{q_0,\tau,\overline{v}}~|~v\in V\}
%$$
From $\Phi$, we build a constraint $\varphi \wedge \psi$ where $\varphi$ is the formula obtained from $\Phi$ by replacing each positive literal $v$ by  $\#v \geq 1$ and each negative literal $\neg v$ by $\# \overline{v} \geq 1$ and  $\psi=\bigwedge_{i=1}^k (\#v_i \geq 1 \wedge \# \overline{v_i}=0) \vee (\# v_i=0 \wedge \# \overline{v_i} \geq 1)$. The former constraint is the natural encoding of
the input propositional formula whereas the latter assigns a consistent interpretation to the control state labels $v_i$ and $\overline{v_i}$
%as assignments to the propositional variable $v_i$. The constraint $\varphi \wedge \psi$ is a formula in \RQNeg. 
as assignments to the propositional variable $v_i$. The constraint $\varphi \wedge \psi$ is a formula in \RQ. 

A node in the initial state $q_0$ makes a guess for the boolean valuation of
a variable $v$ by moving to state $v$ [resp. to $\overline{v}$] if the associated chosen value is $\mathit{true}$ [resp. $\mathit{false}$].
The formula $\psi$ ensures that no contradictory valuation is generated by stating that for each variable $v$ in $V$ 
only one type of control state $v$ or $\overline{v}$ is chosen.
%
%
% Arn 170212
%
%The formula $\Phi$ is part of the \RQNeg-query $\Phi\wedge\Psi$,
%where $\Psi$ is a formula that guarantees that for each variable $x$ in $V$ only one
%label $x$ or $\overline{x}$ is chosen,  i.e.,
%$$
%\Psi=\bigwedge_{i=1}^k (\#v_i \geq 1 \wedge \# \overline{v_i}=0) \vee (\# v_i=0 \wedge \# \overline{v_i} \geq 1)
%$$
Assume that the formula $\Phi$ is satisfiable and let $\set{b_1,\ldots,b_k} \in \set{\mathit{true},\mathit{false}}^k$ be an interpretation over the variables $\set{v_1,\ldots,v_k}$ that satisfies it. From an initial configuration $\gamma_0$ with $k$ nodes, it is possible to reach a configuration $\gamma$ such that $\gamma \models \psi$ and for all $1 \leq i \leq k$ if $b_i=true$ then $\gamma' \models \# v_i \geq 1$ else $\gamma' \models \# v_i =0$. $\gamma \models \varphi \wedge \psi$ clearly holds here. 
Vice versa, if there exists a computation that reaches a configuration that satisfies
$\varphi \wedge\psi$, then we have $m\geq k$ nodes whose labels correspond to a consistent
interpretation of the variables in $V$ and which satisfies $\Phi$.
%
% Arn 170212
%
%We have then that if the formula $\Psi$ is satisfiable, there exists a valuation for
%$v_1,\ldots,v_k$ that satisfies it.
%Indeed, since a formula with label $x$ is satisfied only if $x$ occurs in the configuration,
%label $x$ corresponds to the assign $true$ to variable $x$, whereas label $\overline{x}$ corresponds to assign $false$ to $x$.
%
%We then select an initial configuration with $k$ nodes, generate a labeling corresponding to
%the valuation by executing the local transitions. If each node rewrites in the correct evaluation of a 
%distinct variable, then the corresponding reached configuration satisfies $\Phi\wedge\Psi$.
%
%Vice versa, if there exists a computation that reaches a configuration that satisfies
%$\Phi\wedge\Psi$, then we have $m\geq k$ nodes whose labels correspond to a consistent
%evaluation of the variables in $V$ that satisfy $\Psi$.
\qed
\end{proof}
We will now give an algorithm in \np\ to solve \PRP\ for constraints in \RQ.
As for Algorithm \ref{alg:sfr1}, this new algorithm works on sets of control states. 
The algorithm works in two main phases. In a first phase it generates an increasing sequence of sets of control states 
that can be reached in the considered process definition. At each step the algorithm adds  the control states obtained from the application of the process rules to the current set of labels. 
% Gi 19/2
Unlike the Algorithm \ref{alg:sfr1}, this new algorithm does not merge different branches, i.e. application of distinct rules may 
lead to different sequences of sets of control states.
In a second phase the algorithm only removes control states applying again process rules
in order to reach a set of control states that satisfies the given constraint.
%
% Arn 190212
%
%For a process $\pdef=\tuple{Q,\Sigma,R,Q_0}$  and a set $S \subset Q$, we define the operator $\postadd(\pdef,S) \subseteq 2^Q$ as follows: $S' \in \postadd(\pdef,S)$ if and only if the two following conditions are satisfied (i) $S \subseteq S'$ and (ii) for all $q' \in S' \setminus S$, either there exists a rule $\tuple{q,!!a,q'} \in R$ such that $q \in S$ or there two exists rules $\tuple{p,\broadcast{a},p'}$ and $\tuple{q,\receive{a},q'} \in R$ such that $q,p \in S$ and $p' \in S'$. In other words, all the states present in $S' \in \postadd(\pdef,S)$ are either states of $S$ or states obtained from the application of a broadcast rule of $R$ 
%fireable from control states present in $S$. Similarly, we define the $\postdel(\pdef,S) \subseteq 2^Q$ as follows:  
%$S' \in \postdel(\pdef,S)$ if and only if the two following conditions are satisfied  
%(i) $S' \subseteq S$ and 
%(ii) for all $q \in S \setminus S'$, either there exists a rule $\tuple{q,\broadcast{a},q'} \in R$ such that $q' \in S'$ or there exists two rules $\tuple{p,\broadcast{a},p'} , \tuple{q,\receive{a},q'} \in R$ such that $p,p',q' \in S'$.

\begin{algorithm}[htbp] \footnotesize
%\caption{Solving \PRP\ restricted to \RQNeg}
\caption{Solving \PRP\ for constraints in \RQ}
\label{alg:sfr2}
\begin{algorithmic}
%\REQUIRE $\pdef=\tuple{Q,\Sigma,R,Q_0}$ a process and $\varphi$ a constraint over $\pdef$ in \RQNeg
\REQUIRE $\pdef=\tuple{Q,\Sigma,R,Q_0}$ a process and $\varphi$ a constraint over $\pdef$ in \RQ
\ENSURE  Does $\pdef \models \Diamond \varphi$ ?
\STATE Guess $S_0,\ldots,S_m,T_1,\dots,T_n \subseteq Q$ with $m,n\leq |Q|$
\STATE If $S_0 \not\subseteq Q_0$ return NO
\FORALL{ $i \in \set{0,\ldots,m-1}$}
\STATE If $S_{i+1} \not\in \postadd(\sbp,S_i)$ return NO
\ENDFOR
\STATE $T_0=S_m$
\FORALL{ $i \in \set{0,\ldots,n-1}$}
\STATE If $T_{i+1} \not\in \postdel(\sbp,T_i)$ return NO
\ENDFOR
\STATE If $T_n$ satisfies $\varphi$  return YES else return NO
\end{algorithmic}
\end{algorithm}

For a process $\pdef=\tuple{Q,\Sigma,R,Q_0}$  and a set $S \subset Q$, we define the operator $\postadd(\pdef,S) \subseteq 2^Q$ as follows: $S' \in \postadd(\pdef,S)$ if and only if the two following conditions are satisfied: 
(i) $S \subseteq S'$ and 
(ii) for all $q' \in S' \setminus S$, 
there exists a rule $\tuple{q,!!a,q'} \in R$ such that $q \in S$ 
($q'$ is produced by a broadcast)
or 
there exist rules $\tuple{p,\broadcast{a},p'}$ and $\tuple{q,\receive{a},q'} \in R$ such that $q,p \in S$ and $p' \in S'$
($q'$ is produced by a reception). 
In other words, all the states in $S' \in \postadd(\pdef,S)$ are either in $S$ or states obtained from the 
application of broadcast/reception rules to labels in $S$.
% (we assume here to work in a fully connected graph).
Similarly, we define the operator $\postdel(\pdef,S) \subseteq 2^Q$ as follows:  
$S' \in \postdel(\pdef,S)$ if and only if $S' \subseteq S$ and one of the following conditions hold: either $S \setminus S'=\emptyset$ or [$S \setminus S'=\set{q}$ and there exists a rule $\tuple{q,\broadcast{a},q'} \in R$ such that $q' \in S'$] or [$S \setminus S'=\set{q}$ and there exist two rules $\tuple{p,\broadcast{a},p'}, \tuple{q,\receive{a},q'} \in R$ such that $p,p',q' \in S'$ ($q$ is consumed by a broadcast)] or [$S \setminus S'=\set{p,q}$  and there exist two rules $\tuple{p,\broadcast{a},p'}, \tuple{q,\receive{a},q'} \in R$ such that $p',q' \in S'$ ($p$ and $q$ are consumed by a broadcast)].

%
% Arn 200212
%
%Similarly, we define the operator $\postdel(\pdef,S) \subseteq 2^Q$ as follows:  
%$S' \in \postdel(\pdef,S)$ if and only if the two following conditions are satisfied  
%(i) $S' \subseteq S$ and 
%(ii) for all $q \in S \setminus S'$, either 
%there exists a rule $\tuple{q,\broadcast{a},q'} \in R$ such that $q' \in S'$  
%($q$ is consumed by a broadcast) 
%or 
%there exist rules $\tuple{p,\broadcast{a},p'}, \tuple{q,\receive{a},q'} \in R$ such that $p \in S$ and $p',q' \in S'$  ($q$ is consumed by reception and $p$ can also be consumed).

Finally, we say that a set $S \subseteq Q$ satisfies an atom $\#q=0$ if $q\not\in S$ and it satisfies 
%an atom $\# q\geq 1$ if $q\in S$; satisfiability for composite boolean formulae of \RQ\  is then defined in the natural way.
an atom $\# q\geq 1$ if $q\in S$; satisfiability for composite boolean formulae of \RQ\  is then defined in the natural way.
We have then the following Lemma whose proof can be found in Appendix.

\begin{lemma}\label{theorem:npcomplete}
There is an execution of Algorithm  \ref{alg:sfr2} which answers YES on input $\sbp$ and $\varphi$ iff $\sbp \models \Diamond \varphi$.
%The Algorithm \ref{alg:sfr2} guesses a path in which $\varphi$ holds
%iff there exists an initial configuration $\gamma_0$ from which we can reach  
%a configuration that satisfies $\varphi$.
\end{lemma}
\begin{proof}
Let $\pdef=\tuple{Q,\Sigma,R,Q_0}$ a process with $\ahn(\sbp)=\tuple{\confs,\inttrel,\confszero}$ and $\varphi$ a constraint over $\pdef$ in \RQ.
First we assume that the Algorithm \ref{alg:sfr2} answers YES on input $\sbp$ and $\varphi$. This means that there exists $S_0,\ldots,S_m,T_0,T_1,\dots,\linebreak[0]T_n$ such that $1 \leq m,n \leq |Q|$ and $S_0 \subseteq Q_0$, and for all $i \in \set{0,\ldots,m-1}$, $S_{i+1} \in \postadd(\pdef,S_i)$ and $T_0=S_m$ and for all $i \in \set{0,\ldots,n-1}$, $T_{i+1} \in \postdel(\pdef,T_i)$. We will now prove that there exists two configurations $\gamma_0 \in \confszero$ and $\gamma \in \confs$ such that $\gamma_0 \inttrel^\ast \gamma$ and $L(\gamma)=T_n$. First, as reasoning the same way we did in the proof of Lemma \ref{lemma:algo}, we can deduce that for any $k \in \nat\setminus \set{0}$, there exists $\gamma_0 \in \confszero$ and a complete graph $\gamma_k=\tuple{V,E,L}$ in $\confs $ such that $L(\gamma_k)=S_m$ and for every $q \in S_m$ the set $\set{v \in V \mid L(v) =q}$ has more than $k$ elements. Now we are going to prove that for any $j \in \set{0,\ldots,n}$, for all $k \in \nat \setminus \set{0}$, there is a complete graph $\gamma_{j,k}$ such that:
\begin{enumerate}
\item $L(\gamma_{j,k})=T_j$ and for each $q \in S_j$, the set $\set{v \in V \mid \state(v)=q}$ has more than $k$ elements (i.e. for each element $q$ of $S_j$ there are more than $k$ nodes in $\gamma_{j,k}$ labelled with $q$),
\item there exits $\gamma_0 \in \confszero$ such that $\gamma_0 \inttrel^\ast \gamma_{j,k}$.
\end{enumerate}
To prove this statement we reason by induction on $j$. For $j=0$, since the statement holds for $S_m$, it holds also for $T_0=S_m$. We now assume that the property is true for all naturals smaller than $j$ (with $j <n$) and we will show it is true for $j+1$. We consider now the set $T_j \setminus T_{j+1}$ (assuming it is not empty, otherwise the property trivially holds). By property of the operator $\postdel$, we have $T_{j+1} \subseteq T_j$. Now let $k \in \nat$, the graph $\gamma_{j+1,k}$ is obtained from  $\gamma_{j,k+1}$ as follows: 
\begin{itemize}
\item if $T_{j+1} \setminus T_j=\set{q}$ and there exists a rule $\tuple{q,\broadcast{a},q'} \in R$ such that $q' \in T_{j+1}$], then this rule is applied to all the nodes labelled by $q$; first each node is isolated  with the movement rule, then the broadcast rule is performed and then the complete graph is rebuilt. Note that the application of this rule consecutively will only  increase the number of nodes labelled by $q'$ which were already present in $\gamma_{j,N}$;
\item if $T_{j+1} \setminus T_j=\set{q}$ and there exist two rules $\tuple{p,\broadcast{a},p'}, \tuple{q,\receive{a},q'} \in R$ such that $p,p',q' \in T_{j+1}$ ($q$ is consumed by a broadcast), then all the nodes labelled by $q$ are isolated together with a node labelled by $p$ so that all these nodes are connected, then $p$ broadcast $a$ sending all the other nodes in $q'$ and finally the complete graph is rebuilt; as a consequence there is no more nodes labelled by $q$, the number of nodes labelled by $q'$ and $p'$ have increased and the number of nodes labelled by $p$ has decreased of one unit;
\item if $T_{j+1} \setminus T_j=\set{p,q}$  and there exist two rules $\tuple{p,\broadcast{a},p'}, \tuple{q,\receive{a},q'} \in R$ such that $p',q' \in T_{j+1}$ ($p$ and $q$ are consumed by a broadcast), then as for the second case, we first eliminate all the nodes labelled by $q$ by isolating them together with one node labelled by $p$, and then all the nodes labelled by $p$ can be eliminated the same way it is done in the first case we considered.
\end{itemize}
By applying these rules it is then clear that $\gamma_{j,k+1} \inttrel^\ast \gamma_{j+1,k}$ and also that $\gamma_{j+1,k}$ verifies the property $1$ of the
statement. Since by induction hypothesis, we have that there exists $\gamma_0 \in \confszero$ such that $\gamma_0 \inttrel^\ast \gamma_{j,k+1}$, we also deduce that $\gamma_0 \inttrel^\ast \gamma_{j+1,k}$, hence the property 2 of the statement also holds. Hence if the Algorithm \ref{alg:sfr2} returns YES on input $\pdef$ and $\varphi$, we deduce that there exist a reachable configuration $\gamma \in \confs$ such that $L(\gamma)=T_n$ and  since $T_n$ satisfies $\varphi$, we also have that $\gamma \models \varphi$, hence $\pdef \models \Diamond \varphi$.

%
% Arn 190212
%
%By construction, if $\varphi$ is satisfied in $T_n$, then there exists an initial configuration
%from which we can reach configuration with labels in $T_n$ that satisfies $\varphi$.
%Indeed $post^1$ mimicks an execution in which labels that may disappear are moved away from the 
%sender node, whereas $post^2$ mimicks an execution that involves only a sender and a receiver 
%whose label is cancelled (all other nodes move away from the sender).
%
We now assume that there exists two configurations $\gamma_0 \in \confszero$ and $\gamma \in \confs$ such that $\gamma_0 \inttrel^+ \gamma$ (the case $\gamma_0=\gamma$ can be easily verified) and $\gamma \models \varphi$. Hence there exists $\gamma_1,\ldots,\gamma_k \in \confs$ such that $\gamma_0 \inttrel^+ \gamma_1 \ldots \inttrel^+ \gamma_k$ with $\gamma_k=\gamma$ and for all $i \in \set{1,\ldots,k}$, exactly one broadcast rule has been applied between $\gamma_i$ and $\gamma_{i+1}$. From this execution we build a sequence of set of control states $(S'_i)_{0 \leq i \leq k}$ such that $S'_0=L(\gamma_0)$ and for all $0 \leq i \leq k-1$, $S'_{i+1}=S'_i \cup L(\gamma_i)$ . By definition of the broadcast rule and of the operator $\postadd$, we deduce that $S'_{i+1} \in \postadd(\pdef,S'_i)$. From this sequence, we can furthermore extract a subsequence $(S_i)_{0 \leq i \leq m}$ such that for all $0 \leq i \leq m-1$, $S_{i+1} \in \postadd(\pdef,S_i)$ and $S_{i+1} \neq S_i$ and for all $0 \leq j \leq k$, there exists $0 \leq i \leq m$ such that $S'_j=S_i$. Since we have $S_{i} \subset S_{i+1}$ for all $0 \leq i \leq m-1$, we deduce that necessarily $m \leq |Q|$.  Now we build another sequence of control states $(T'_i)_{0 \leq i \leq k}$ such that $T'_0=S_m$ and for all $0 \leq i \leq k-1$, $T'_{i+1}=T'_{i} \setminus E_i$ where for all $0 \leq i \leq k-1$, $E_i=\set{ q \in L(\gamma_i) \mid \not\exists j > i \mbox{ s.t. } q \in L(\gamma_j)}$.  In other words, to build $T'_{i+1}$ from $T'_i$ we delete the control states $q$ that are present in $\gamma_i$ and will never be present in any $\gamma_j$ for $j >i$. We recall that by construction for all $1 \leq i \leq k$, we have $L(\gamma_i) \subseteq T'_0$ and hence by construction of the sequence $(T'_i)_{0 \leq i \leq k}$ we have necessarily $L(\gamma)=T'_k$. By definition of the broadcast rule and of the operator $\postdel$, we also deduce that $T'_{i+1} \in \postdel(\pdef,T'_i)$. From this sequence, we can furthermore extract a subsequence $(T_i)_{0 \leq i \leq n}$ such that for all $0 \leq i \leq n-1$, $T_{i+1} \in \postdel(\pdef,T_i)$ and $T_{i+1} \neq T_i$ and for all $0 \leq j \leq k$, there exists $0 \leq i \leq n$ such that $T'_j=T_i$. Since we have $T_{i+1} \subset T_{i}$ for all $0 \leq i \leq n-1$, we deduce that necessarily $n \leq |Q|$ and also we have $T(n)=L(\gamma)$. Since $\gamma \models \varphi$, we deduce that $T_n$ satisfies $\varphi$ and consequently we have proved that there is an execution of Algorithm \ref{alg:sfr2} which answers YES on input $\pdef$ and $\varphi$.
%
% Arn 190212
%
%Assume that there exists an initial configuration $\gamma_0$ from which we can reach a
%configuration $\gamma'$ that satisfies $\varphi$.
%Let $S$ be the set of labels occurring in $\gamma'$ and let $\{q_1,\ldots,q_k\}$ be its complement.
%Furthermore, let $r_1,\ldots,r_k$ be the sequence of rules that consume
%the last occurrence of resp. $q_1,\ldots,q_k$ in the derivation from $\gamma_0$ to $\gamma'$.
%By construction, there exists a path of length at most $|Q|$
%going from $L(\gamma_0)$ to the set $S'$ that contains all labels occurring
%in the derivation from $\gamma_0$ to $\gamma'$.
%Starting from $S'$ we can now apply in sequence $r_1,\ldots,r_k$ so as to remove
%one by one the states $q_1,\ldots,q_k$.
%Indeed, in $S'$ we have all preconditions needed to fire $r_1$.
%We fire it so as to consume only $q_1$ (this is possible by construction of $post^2$).
%So now in $S'\setminus\{q_1\}$ we have all preconditions needed to fire $r_2$, we fire
%$r_2$ so as to consume only $q_2$, and so on.
%The resulting set of labels can then be guessed non-deterministically by our algorithm.
\qed

\end{proof}
It is then clear that each check performed by the Algorithm \ref{alg:sfr2} (i.e. $S_0 \subseteq Q_0$ and $S_{i+1} \in \postadd(\pdef,S_i)$ and $T_{i+1} \in \postadd(\pdef,T_i)$ and $T_n$ satisfies $\varphi$) can be performed in polynomial time in the size of the process $\pdef$ and of the formula $\varphi$ and since $m$ and $n$ are smaller than the number of control states in $\pdef$, we deduce the following theorem (the lower bound being given by Proposition \ref{theorem:nphard}).
\begin{theorem}
%\label{theorem:npcomplete}
%\PRP\ restricted to \RQNeg\ is \np-complete.
\PRP\ for constraints in \RQ\ is \np-complete.
\end{theorem}

%% file: full_queries_petri.tex
\newcommand{\MCon}{{\cal M}}
\newcommand{\denote}[1]{[\!\![{#1}]\!\!]}

%\section{Complexity for \PRP\ }
%
\section{Complexity of the Cardinality Reachability Problem}
In this section we study another problem, we call CRP, in which we ask the question 
whether we can reach a configuration with a given number of occurrences for each control state.
%Given a number of occurrences for each of the control state of a process, we want to check if it is %possible to reach a configuration where each control state appears exactly this number of times.
Formally, given a process $\sbp=\tuple{Q,\Sigma,R,Q_0}$, a \emph{cardinality constraint} over $\sbp$ is a function $\mathit{card} : Q \rightarrow \nat$. We say that a configuration $\conf$ satisfies a cardinality constraint $\mathit{card}$ (denoted by $\conf \vdash \mathit{card}$) if for each $q \in Q$ the number of occurrences of $q$ in $\conf$ is equal to $\mathit{card}(q)$. The Cardinality Reachability Problem (CRP) can then be stated as follows:
\begin{description}
\item[Input:] A process $\sbp$ with  $\ahn(\sbp)=\tuple{\confs,\inttrel,\confs_0}$ and a cardinality constraint $\mathit{card}$.
%\item[Input:] A process $\sbp=\tuple{Q,\Sigma,R,Q_0}$ with  $\ahn(\sbp)=\tuple{\confs,\Rightarrow,\confs_0}$, 
%              and a query $\varphi$.
\item[Output:] Yes, if $\exists\gamma_0 \in \confs_0$ and $\gamma_1\in\confs$ s.t. $\gamma_0\inttrel^*\gamma_1$ and $\gamma_1 \vdash \mathit{card}$.
\end{description} 
%
%We now consider the parameterized reachability problem.
Note that this problem seems easier than \PRP\ because the cardinality constraint fixes the number of nodes of an initial configuration. In fact, if there is a reachable configuration which satisfies a cardinality constraint $\mathit{card}$, we know that this configuration and the initial configuration from which the computation starts have $\Sigma_{q \in Q} \mathit{card}(q)$ nodes. 
We will show that this is not the case as CRP is \pspace-complete.
% in its full generality.
%i.e., queries in which atoms have the form $\#q \sim k$ for some relation $\sim$.
%By resorting to Petri nets as done in \cite{concur} we could 
%show that \PRP\ is decidable by 
%reduction to the Petri net reachability problem,
%which is know to require exponential space~\cite{lipton76}.
%% (actually we can use target reachability as in \cite{BZ01}).
%In this section we show that it is not necessary to resort to the full power of Petri nets as we can give dedicated symbolic algorithms for solving the problem using polynomial space.
%%
%
First we prove the lower bound.
\begin{proposition}
\label{prop:pspacehard}
CRP\ is \pspace-hard.
\end{proposition}
\begin{proof}
We use a reduction from reachability in 1-safe Petri nets.
A Petri net $N$ is a tuple $N=\tuple{P,T,\vec{m_0}}$, where $P$ is a finite set of places, $T$ is a finite set of transitions $t$, such that $^\bullet t$ and $t^\bullet$ are multisets of places (pre- and post-conditions of $t$),
and $\vec{m_0}$ is a multiset of places
%marking, i.e., a function from places to natural numbers 
that indicates how many tokens
are located in each place in the initial net marking. 
%We use $\vec{m}\sim\vec{m'}$ to denote pointwise operations and comparisons 
%on vectors ($\sim\in\{+,-,\geq\}$).
Given a marking $\vec{m}$, the firing of a transition $t$ such that $^\bullet t\subseteq \vec{m}$ 
leads to a new marking $\vec{m'}$ obtained as $\vec{m'}=\vec{m}\setminus^\bullet t\cup t^\bullet$.
A Petri net $P$ is 1-safe if in every reachable marking every place has at most one token.
Reachability of a specific marking $\vec{m_1}$ from the initial marking $\vec{m_0}$ is decidable for Petri nets,
and \pspace-complete for 1-safe nets \cite{onesafe}.

Given a 1-safe net $N=\tuple{P,T,\vec{m_0}}$ and a marking $\vec{m_1}$, we encode the reachability problem 
as a CRP problem for the process $\pdef$ and cardinality constraint 
$\mathit{card}$
defined next.
For each place $p\in P$, we introduce control states $p_{1}$ and $p_{0}$
to denote the presence or absence of the token in $p$, respectively. 
Furthermore, we introduce a special control state $ok$.
The control state is used to control the net simulation.
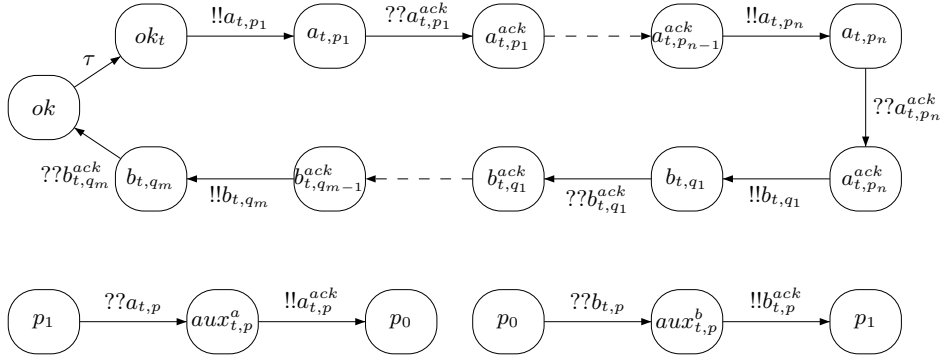
\begin{figure}[t]
\begin{center} 
\scalebox{.95}{
\begin{picture}(125,40)(0,-40)
  \node[Nw=10,Nh=9](OK)(0,-10){$ok$}
	\node[Nw=10,Nh=9](A)(15,0){$ok_t$}
  \node[Nw=10,Nh=9](B)(40,0){$a_{t,p_1}$}
  \node[Nw=10,Nh=9](C)(65,0){$a^{ack}_{t,p_1}$}
  \node[Nw=10,Nh=9](D)(90,0){$a^{ack}_{t,p_{n-1}}$}
  \node[Nw=10,Nh=9](E)(115,0){$a_{t,p_{n}}$}
  \node[Nw=10,Nh=9](F)(115,-20){$a^{ack}_{t,p_{n}}$}
  \node[Nw=10,Nh=9](G)(90,-20){$b_{t,q_1}$}
  \node[Nw=10,Nh=9](H)(65,-20){$b^{ack}_{t,q_1}$}
  \node[Nw=10,Nh=9](I)(40,-20){$b^{ack}_{t,q_{m-1}}$}
  \node[Nw=10,Nh=9](J)(15,-20){$b_{t,q_{m}}$}
  \drawedge[ELside=l](OK,A){$\tau$}
  \drawedge[ELside=l](A,B){$\broadcast{a_{t,p_1}}$}
  \drawedge[ELside=l](B,C){$\receive{a^{ack}_{t,p_1}}$}
  \drawedge[dash={1.5}0](C,D){}
  \drawedge[ELside=l](D,E){$\broadcast{a_{t,p_{n}}}$}
  \drawedge[ELside=l](E,F){$\receive{a^{ack}_{t,p_{n}}}$}
  \drawedge[ELside=l](F,G){$\broadcast{b_{t,q_1}}$}
  \drawedge[ELside=l](G,H){$\receive{b^{ack}_{t,q_1}}$}
  \drawedge[dash={1.5}0](H,I){}
  \drawedge[ELside=l](I,J){$\broadcast{b_{t,q_{m}}}$}
  \drawedge[ELside=l](J,OK){$\receive{b^{ack}_{t,q_{m}}}$}
  \node[Nw=10,Nh=9](K)(0,-40){$p_{1}$}
  \node[Nw=10,Nh=9](L)(25,-40){$aux^{a}_{t,p}$}
  \node[Nw=10,Nh=9](M)(50,-40){$p_{0}$}
  \drawedge(K,L){$\receive{a_{t,p}}$}
  \drawedge(L,M){$\broadcast{a^{ack}_{t,p}}$}
  \node[Nw=10,Nh=9](N)(65,-40){$p_{0}$}
  \node[Nw=10,Nh=9](O)(90,-40){$aux^{b}_{t,p}$}
  \node[Nw=10,Nh=9](P)(115,-40){$p_{1}$}
  \drawedge(N,O){$\receive{b_{t,p}}$}
  \drawedge(O,P){$\broadcast{b^{ack}_{t,p}}$}
%
  %\node[Nw=8,Nh=8](P)(100,-40){$p_{0}$}
  %\node[Nw=8,Nh=8](Q)(120,-40){$aux_{p}$}
  %\node[Nw=8,Nh=8](R)(140,-40){$p_{1}$}
  %\drawedge(P,Q){$\receive{c_{t,p}}$}
  %\drawedge(Q,R){$\broadcast{ack_{t}}$}

%  \node[Nw=8,Nh=8](OK)(0,0){$ok$}
%  \node[Nw=8,Nh=8](A)(0,-20){$ok_t$}
%  \node[Nw=8,Nh=8](B)(30,-20){$a_{1}$}
%  \node[Nw=8,Nh=8](C)(50,-20){$a_{n-1}$}
%  \node[Nw=8,Nh=8](D)(80,-20){$a_{n}$}
%  \node[Nw=8,Nh=8](E)(110,-20){$b_{1}$}
%  \node[Nw=8,Nh=8](R)(130,-20){$b_{m-2}$}
%  \node[Nw=8,Nh=8](F)(160,-20){$b_{m-1}$}
%  \node[Nw=8,Nh=8](G)(130,0){$b_{m}$}
%  \drawedge[ELside=r](OK,A){$\tau$}
%  \drawedge[ELside=r](A,B){$\broadcast{a_{t,p_1}}$}
%  \drawedge[ELside=r](B,C){$\ldots$}
%  \drawedge[ELside=r](C,D){$\broadcast{a_{t,p_{n}}}$}
%  \drawedge[ELside=r](D,E){$\broadcast{b_{t,q_1}}$}
%  \drawedge[dash={1.5}0](E,R){$\ldots$}
%  \drawedge[ELside=r](R,F){$\broadcast{b_{t,q_{m-1}}}$}
%  \drawedge[ELside=r](F,G){$\broadcast{c_{t,q_{m}}}$}
%  \drawedge[ELside=r](G,OK){$\receive{ack_t}$}

%  \node[Nw=8,Nh=8](H)(0,-40){$p_{1}$}
%  \node[Nw=8,Nh=8](I)(20,-40){$p_{0}$}
%  \drawedge(H,I){$\receive{a_{t,p}}$}
  
%  \node[Nw=8,Nh=8](L)(60,-40){$p_{0}$}
%  \node[Nw=8,Nh=8](M)(80,-40){$p_{1}$}
%  \drawedge(L,M){$\receive{b_{t,p}}$}

%  \node[Nw=8,Nh=8](P)(100,-40){$p_{0}$}
%  \node[Nw=8,Nh=8](Q)(120,-40){$aux_{p}$}
%  \node[Nw=8,Nh=8](R)(140,-40){$p_{1}$}
%  \drawedge(P,Q){$\receive{c_{t,p}}$}
%  \drawedge(Q,R){$\broadcast{ack_{t}}$}
%
\end{picture}
}
\end{center}
\caption{Simulation of a transition $t$ with $^\bullet t=\{p_1,\ldots,p_n\}$ and $t^\bullet=\{q_1,\ldots,q_m\}$.
}
\label{onesafe}
\end{figure}
Transitions of the controller are depicted in the upper part of Fig. \ref{onesafe}. 
The first rule of the controller selects the current transition to simulate.
The simulation of the transition $t$ with $^\bullet t=\{p_1,\ldots,p_n\}$ and $t^\bullet=\{q_1,\ldots,q_m\}$ is defined via two sequences of messages.
The first one is used to remove the token from 
$p_1,\ldots,p_n$, whereas the second one
is used to put the token in $q_1,\ldots,q_m$.
To guarantee that every involved place reacts to the protocol ---i.e. messages are not lost--- the controller waits for an acknowledgment from each of them. Transitions of places are depicted in the lower part of Fig. \ref{onesafe}.
It is not restrictive to assume that there is only one
token in the initial marking $\vec{m_0}$
(otherwise we add an auxiliary initial place
and a transition that generates $\vec{m_0}$
by consuming the initial token).
Let $p^0$ be such a place.
We define the initial states $Q_0$
of the process $\pdef$ as $\{p^0_1,ok\}\cup\{p_0\ |\ p\in P\setminus\{p^0\}\}$,
in order to initially admit control states
representing the controller, the presence of the initial token,
and the absence of tokens in other places.
%the initial set of control states $Q_0$, 
%i.e., $Q_0$ contains the state $p_{1}$ for each $p$ occurring in $\vec{m_0}$, and $q_{0}$ for each $q$ not in $\vec{m_0}$.
%Furthermore, it also contains the controller state $ok$.
%The final marking $\vec{m_1}$ is encoded via a constraint $\varphi$
%in which we require that 
%$\#p_{1}= 1$ for each $p$ occurring in $\vec{m_1}$
%and $\#q_{1}= 0$ for the other places
%(i.e. $\varphi = [\wedge_{p\in \vec{m_1}} (\#p_{1}= 1)
%\wedge \wedge_{q\not\in \vec{m_1}} (\#q_{1}= 0)]$).
The reduction does not work if there are several copies of controller nodes and/or
place representations (i.e. $p_{1},p_{0},\ldots$) interacting during a simulation (interferences
between distinct nodes representing controllers/places may lead to incorrect results).
However we can ensure that the reduction is accurate by checking
the number of occurrences of states exposed in the final
configuration: it is sufficient to check that only one
controller and only one node per place
in the net are present. Besides making this check,
the cardinality constraint $\mathit{card}$
should also verify that the 
represented net marking coincides with $\vec{m_1}$.
Namely, we define $\mathit{card}$ as follows:
$$
\begin{array}{l}
\forall {p \in \vec{m_1},t \in T}. 
\left (
\begin{array}{l}
\mathit{card}(p_{1})= 1 \
\wedge\ \mathit{card}(p_{0})= 0\ \wedge \\
\mathit{card}(aux^a_{t,p})=0\ \wedge\ \mathit{card}(aux^b_{t,p})=0
\end{array}
\right )
\ \wedge \\
\forall {q \not\in \vec{m_1},t \in T}. 
\left (
\begin{array}{l}
\mathit{card}(q_{1})= 0 \
\wedge\ \mathit{card}(q_{0})= 1\ \wedge \\
\mathit{card}(aux^a_{t,q})=0\ \wedge\ \mathit{card}(aux^b_{t,q})=0
\end{array}
\right )
\ \wedge \\
\mathit{card}(ok)= 1\ \wedge\ \forall t\in T.\mathit{card}(ok_t)= 0\ \wedge
\\
\forall_{t\in T,q\in P} \big(\mathit{card}(a_{t,q})=0 \wedge \mathit{card}(b_{t,q})=0 \wedge \mathit{card}(a^{ack}_{t,q})=0 \wedge \mathit{card}(b^{ack}_{t,q})=0 \big)
\end{array}
$$
%adding auxiliary conditions to the query.
%Let $\otimes$ be the logical operator exclusive or.
%We define the condition $\psi=[\#ok=1 \wedge (\wedge_{t \in T} \#ok_t=0) \wedge \psi_1\wedge\psi_2]$,
%where $\psi_1=\bigwedge_{p\in P}   [(\#p_{1}=1 \bigotimes \#p_{0}=1) \wedge \big(
%\wedge_{t \in T} (\#aux^a_{t,p}=0 \wedge \#aux^b_{t,p}=0)\big)]$ 
%and  $\psi_2=\bigwedge_{t\in T,q\in P} [\#a_{t,q}=0 \wedge \#b_{t,q}=0 \wedge \#a^{ack}_{t,q}=0 \wedge \#b^{ack}_{t,q}=0]$.
%The constraint $\Psi$ is defined as $\varphi\wedge\psi$.
Since the number of nodes stays constant during an execution, the post-condition specified by $\mathit{card}$ is propagated back to the initial configuration.
Therefore, if the protocol satisfies CRP for $\mathit{card}$, then 
in the initial configuration there must be one single controller node with state $ok$, and for each place $p$ %in the initial marking 
one single node with either state
$p_{1}$ or state $p_0$. 
%For all other place $q$, we need a single node with state $q_{0}$.
%Finally, there cannot be nodes with other types of labels.
Under this assumption, it is easy to check that a run of the protocol corresponds precisely to a firing sequence in the 1-safe net. Thus an execution run satisfies $\mathit{card}$ if and only if the corresponding firing sequence reaches the marking $\vec{m_1}$.
\hfill\qed
\end{proof}

%% file: reachalg.tex
We now show that there exists an \npspace\ algorithm to decide CRP.
Let $\pdef=\tuple{Q,\Sigma,R,Q_0}$. 
Since the size of a graph never changes during an execution, a cardinality constraint 
fixes the size of the initial configuration given by the sum $K$ of constants in $\mathit{card}$.
The algorithm guesses an execution 
$\gamma_0\inttrel\gamma_1\inttrel\ldots\inttrel\gamma_n$ 
traversing pairwise distinct configurations,
s.t. $\gamma_0$ is a complete graph with $K$ nodes 
in initial states, and then checks if $\mathit{card}$ is satisfied in $\gamma_n$.
Each configuration can be stored in polynomial space.
Since the size of all configurations is $K$ 
we need at most $K^{|Q|}$ (all possible combinations of states over $K$ nodes). 
%We can keep count the length of a path using polynomial space.
Thus we have a non-deterministic algorithm working in polynomial space.
Since \npspace=\pspace, and in the light of the lower bound
indicated by Proposition~\ref{prop:pspacehard},
we can conclude with the following theorem.
\begin{theorem}
CRP is \pspace-complete.
\end{theorem}

%% file: concl.tex
\section{Conclusion} 
We have studied the complexity of reachability problems for mobile ad hoc network
protocols in which 
target states are represented by using constraints 
checking the presence, absence, or counting the number of
occurrences of control states in a configuration.
We have given algorithms for different classes of constraints.
For constraints that simply checks the presence of control states
we have shown that reachability is \ptime-complete, while when
also constraints checking 
the absence are considered the problem turns out to be \np-complete.
Finally, 
for constraints counting the number of occurrences reachability
becomes \pspace-complete. 
%As for lossiness in other computational models, spontaneous movement simplifies the verification task
%for problems like coverability.
Our analysis significantly improves  
the decidability results given in \cite{concur}
by reduction to problems which are known to be 
at least \expspace-hard.